# Skyrmion Bag Robustness in Plasmonic Bilayer and Trilayer Moiré Superlattices

Julian Schwab[1*], Florian Mangold[1], Bettina Frank[1], Timothy J. Davis[1,2,3], Harald Giessen[1]

[1]*4th Physics Institute, Research Center SCoPE, and Integrated Quantum Science and Technology Center, University of Stuttgart; 70569 Stuttgart, Germany.*

[2]*Faculty of Physics and Center for Nanointegration, Duisburg-Essen (CENIDE). University of Duisburg-Essen; 47048 Duisburg, Germany.*

[3]*School of Physics, University of Melbourne; Parkville Victoria 3010, Australia.*

*\*Corresponding author. Email: julian.schwab@pi4.uni-stuttgart.de*

**Abstract**

Twistronics is studied intensively in twisted 2D heterostructures and its extension to trilayer moiré structures has proven beneficial for the tunability of unconventional correlated states and superconductivity in twisted trilayer graphene. Just recently, the concept of twistronics has been applied to plasmonic lattices with nontrivial topology, demonstrating that bilayer moiré skyrmion lattices harbor multi-skyrmion textures called skyrmion bags. Here, we explore the properties of plasmonic trilayer moiré superlattices that are created by the interference of three twisted skyrmion lattices. More specifically, we explore the properties of periodic superlattices and their topological invariants. We also demonstrate that twisted trilayer skyrmion lattices harbor the same skyrmion bags as twisted bilayer skyrmion lattices. We quantify the robustness of these skyrmion bags by the stability of their topological numbers against certain disturbance fields that leads to experimental designs for topological textures with maximum robustness.

**Main Text**

Moiré patterns in atomic systems can be constructed by stacked layers of van der Waals materials with a relative twist and/or lattice mismatch. The moiré pattern modulates the lattice periodically resulting in a superlattice demonstrating a variety of strongly correlated phases. These superlattices were studied intensively in twisted bilayer graphene, where this enables superconductivity [1], correlated insulating states [2], and strange metal behavior [3]. Recently, huge attention has also been drawn towards trilayer moiré superlattices (TMSLs) such as twisted trilayer graphene [4–9], hexagonal boron nitride (h-BN)-graphene heterostructures [10–12], and transition metal dichalcogenide heterostructures [13–15]. The introduction of a third layer modulates the bilayer moiré superlattice and leads to long-range super-moiré patterns that are controlled by the twist angle of the third layer. This new twist angle adds more degrees of freedom and offers more flexibility, e.g., to tune unconventional correlated states and superconductivity in twisted trilayer graphene [4, 8, 9, 16].

In optics, twisted bilayer photonic lattices have been created using metasurfaces, metamaterials, and photonic crystals [17–23]. However, trilayer photonic lattices have so far only been realized to study the far field coupling of a moiré nano particle lattice laser [24] with many remaining aspects to be investigated. Here we focus on the topology of trilayer twisted optical lattices with nontrivial topological features such as skyrmions [25]. These are three-dimensional topological defects on a two-dimensional plane that were observed first in solid materials [26–29] and liquid crystals [30, 31]. Skyrmions were also demonstrated in optics [32–35] and particularly in plasmonic systems [36–38], where skyrmion lattices emerge through interference of surface plasmon polariton (SPP) waves. Recently, the concept of twistronics has been applied to such plasmonic lattices in order to create plasmonic moiré skyrmion superlattices [39]. By investigating their topology, it was demonstrated that they harbor skyrmion bags [28, 31, 40], which are multi-skyrmion textures that consist of $N$ skyrmions contained within a skyrmion boundary of opposite winding number $-1$. Their size is controllable by the twist angle and the center of rotation, and they feature a degree of robustness against deviations of the twist angle [39].

Here, we extend the concept of 'plasmonic twistronics' to trilayer plasmonic skyrmion superlattices, consisting of three twisted plasmonic skyrmion lattices. We present numerical results of commensurate super-moiré lattices with large periodicity and a super-cell with very large topological charge. We discover that they feature skyrmion bags of new shapes that are not possible to realize in twisted bilayer skyrmion lattices. In addition to these new bag textures, all skyrmion bags of the bilayer system can be recreated using mirror-symmetric trilayer skyrmion lattices. We demonstrated previously that bilayer skyrmion bags possess a degree of topological robustness against deviations of the twist angle [39]. This robustness increases for smaller skyrmion bags that appear at larger twist angles.

In the following, we present a method to quantify the topological robustness of skyrmion bags by testing their stability against perturbational fields. This allows us to compare the skyrmion bag robustness in bilayer and trilayer moiré superlattices and determine the ideal twist angles in these systems. Such a comparison of the topological stability is helpful when designing the ideal structures for experimental realizations. We demonstrate that this general approach can also be used to compare the robustness of different topological textures and assess their chance of a successful experimental implementation. Our results extend the toolbox of plasmonic twistronics and provide a valuable method to determine topological textures in wave systems with high topological stability. These textures might find subsequent application in structured light-matter interactions.

**Plasmonic Trilayer Moiré Skyrmion Superlattices**

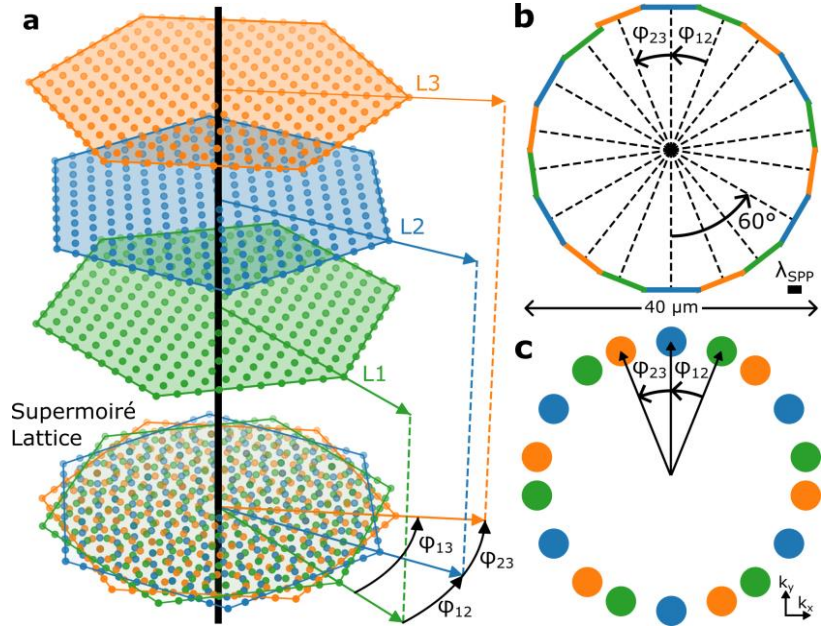

**Figure 1 | Plasmonic super-moiré skyrmion lattices. a**, Super-moiré lattice created by three hexagonal lattices with relative twist $\phi_{12}$ between the first and second lattice (L1 and L2) and $\phi_{23}$ between the second and third lattice (L2 and L3). In plasmonics, the lattices are superimposed in a single layer as illustrated at the bottom. **b**, corresponding coupling structure of a super-moiré skyrmion lattice. The coupling structure consists of three tilted hexagons, generating three skyrmion lattices with relative twists. **c**, schematic Fourier decomposition of a moiré skyrmion superlattice, where the colors originate from the boundaries in b and the lattice colors in a.

Plasmonic lattices can be excited by specific coupling structures that are etched into a single crystalline gold flake using ion beam lithography [36, 41, 42]. A laser pulse incident on such grooves in a metal surface generates charge excitations on their edges, from where SPP waves are launched and propagate towards the center of the structure. The interference pattern of these plasmon waves can exhibit different topological features in their electric field vectors [36, 41] and their spin angular momentum vectors [37, 42–44] that are controlled by the shape of the coupling structure and the resulting propagation direction and phase shift of the SPPs [45]. In the case of a hexagonal coupling structure, a plasmonic skyrmion lattice can be generated [36]. Plasmonic skyrmion lattices exhibit a six-fold symmetry [42], reminiscent of two-dimensional honeycomb lattices in graphene, h-BN, or hexagonal TMDC monolayers. Moiré lattices have been investigated intensively using these van-der-Waals materials, and recent research has also been directed towards twisted trilayer heterostructures, such as twisted trilayer graphene.

Using such materials, TMSLs are created by stacking three layers of a 2D lattice with a relative twist, as illustrated in Fig. 1a. Superimposing, the three layers results in a super-moiré lattice as illustrated at the bottom. The moiré twist angles $\phi_{12}$ ($\phi_{23}$) between the consecutive lattices L1 and L2 (L2 and L3) are connected to the twist between L1 and L3 by $\phi_{13} = \phi_{12} + \phi_{23}$.

In optics, and specifically in plasmonics, moiré lattices are created by the interference of superimposed photonic lattices with a relative twist. Here, we focus on plasmonic trilayer moiré skyrmion lattices that are created by the interference of three twisted skyrmion lattices. The corresponding coupling structure is composed of three twisted hexagons, with each hexagon contributing a skyrmion lattice with different orientation. The three hexagons are combined into a coupling structure which consists of 18 line segments. The coupling structure is illustrated in Fig. 1b, where the three colors of the boundaries illustrate the origin of the boundaries from the three hexagons. The excitation of SPPs from all boundaries with the same efficiency is realized by using circularly polarized light. This creates a phase shift between SPPs that are launched from boundaries with different orientation, as a result of spin-orbit coupling [45]. To regain constructive interference in the center of the structure, this phase shift is compensated by adjusting the positions of the boundaries akin to an Archimedean spiral. The Fourier decomposition of the resulting skyrmion TMSL electric field distribution in Fig. 1c highlights the six-fold symmetry of the three skyrmion lattices and their hexagonal excitation boundaries, as well as the moiré twist angles between them.

Bilayer moiré superlattices possess long-range aperiodic order and are hence quasicrystalline [9, 46] for general twist angles. Nonetheless, periodic bilayer structures can be obtained using specific twist angles that rotate a lattice site of one layer onto a lattice site of the other layer. The corresponding commensurability in six-fold symmetry has been first derived for twisted bilayer graphene [47, 48] and can be denoted by the integers $m$ and $n$ with the corresponding mapping

$$m\boldsymbol{a_1} + n\boldsymbol{a_2} \to n\boldsymbol{a_1} + m\boldsymbol{a_2}, \qquad m, n \in \mathbb{Z}, \tag{1}$$

as illustrated in Fig. 2a, b. These angles have recently also been utilized to create periodic plasmonic bilayer moiré skyrmion superlattices [39]. In the case of three layers, the moiré pattern is defined by setting two out of the three twist angles $\{\phi_{12}, \phi_{23}, \phi_{13}\}$. This effectively creates two bilayer moiré patterns that are combined to a more complex super-moiré pattern. Periodicity of the TMSL is achieved by choosing both individual twist angles to be bilayer commensurate angles that fulfill the moiré mapping equation (1) [5, 49, 50]. With this approach, a family of super-moiré skyrmion lattices are generated and depicted in Fig. 2c-g with their associated superlattice vectors. Mathematical details about the commensurability and the resulting superlattice period are included in the supplement. The electric field distributions are derived assuming infinitely long boundary lines enabling the approximation of the SPP wave as a plane wave. This approximation results in an infinitely large field of view allowing the periodicity to be clearly observed. In practice, the field of view is limited by the size of the structure and the duration of the excitation pulse.

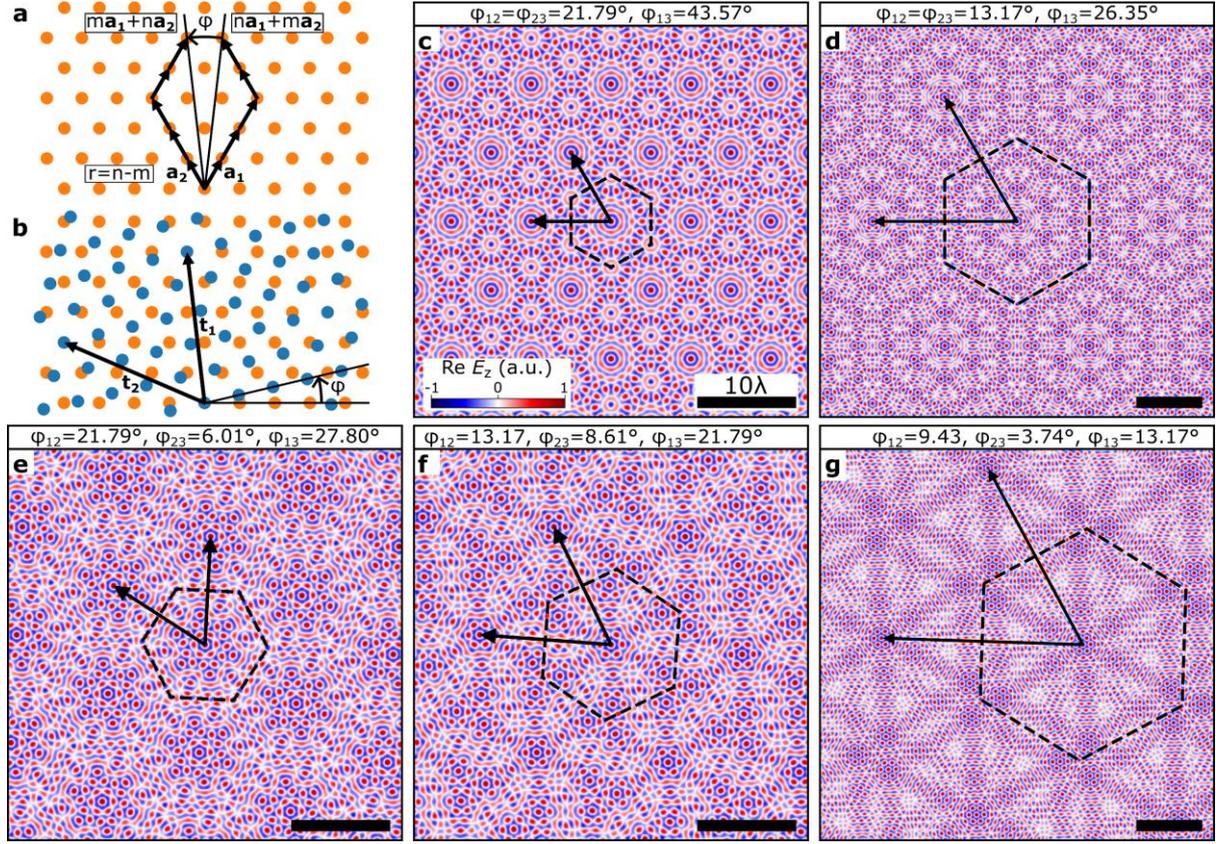

**Figure 2 | Periodic plasmonic super-moiré skyrmion lattices. a**, Periodic bilayer superlattices are created by mapping the lattice site $na_1 + ma_2$ to $ma_1 + na_2$ using the twist angle $\phi$. **b**, Moiré superlattice and superlattice vectors $t_1$ and $t_2$ for the rotation $\phi$ indicated in a. In the case of three twisted lattices commensurability is achieved if the twist angles $\phi_{12}$ and $\phi_{13}$ are both such bilayer commensurate angles. **c**-**g**, Exemplary periodic trilayer moiré skyrmion superlattices and their superlattice vectors. The superlattice unit cells are surrounded by dashed lines.

## Topological Properties of Trilayer Moiré Skyrmion Superlattices

We analyze the local topology of the TMSL vector fields using the skyrmion number density

$$s = \frac{1}{4\pi}\hat{e} \cdot \left(\frac{\partial \hat{e}}{\partial x} \times \frac{\partial \hat{e}}{\partial y}\right), \qquad (2)$$

where $\hat{e} = E(r,t)/||E(r,t)||$ is the unit vector of the SPP electric field. The skyrmion number or topological charge is defined as $S = \int_\sigma s \, dA$, where $\sigma$ is the local region in which the topology is examined. In skyrmion lattices each unit cell consists of a skyrmion with $S = 1$. Analogously, the superlattice unit cells of bilayer [39] and trilayer skyrmion moiré superlattices possess a well-defined topological charge that can become very large for large periodicities of the superlattice. These topological charges are rigorously defined by the moiré superlattice and can therefore be used to characterize its topology. All topological charges of the super unit cells of Fig. 2 are listed in Table 1. These values differ from the total number of skyrmion lattice sites that are contained in the super unit cell. This is because the twisted skyrmion lattices are superimposed in a single plane, which leads to an interference of these lattices. The interference can merge multiple skyrmions of different lattices to single elongated skyrmions. This decreases the total topological charge and gives rise to new topological features in the moiré superlattice such as skyrmion bags. Skyrmion bags have been identified first in liquid crystals and chiral magnets [28, 31, 40] and have recently been discovered in plasmonic twisted bilayer skyrmion lattices. They represent multi-skyrmion configurations, where $N$ skyrmions with $S =$ 1 are enclosed within a larger skyrmion. This surrounding skyrmion has oppositely oriented field components and therefore $S = -1$. Consequently, the total topological charge of a skyrmion bag is $S_{bag} = N - 1$.

In moiré skyrmion lattices, skyrmion bags are created through the interference of overlapping skyrmions. This interference creates a closed loop with positive out-of-plane field component. The size of the loop is dependent on the twist angle and the rotation center and encloses a cluster of $N$ individual skyrmions, thus forming the skyrmion bag. Likewise, the cluster skyrmions are also composed of multiple overlapping skyrmions. However, their displacement is typically not large enough for them to merge with neighboring skyrmions which is why they form elongated skyrmions. The skyrmion number of the cluster can be obtained by integrating the skyrmion number density using an integration boundary that is set along the edge of the non-rotated unit cells (red dashed line of Fig. 3e). For the full skyrmion bag the integration boundary can be set along the positions of the surrounding overlapping skyrmions (black dashed line of Fig. 3e).

These skyrmion bags can be observed in and around the center of the superlattice unit cells of Fig. 2. Additionally, trilayer moiré lattices give rise to skyrmion bag textures which contain other skyrmion bags (see Fig. 2e, f, g) and therefore have new topological charges that may not be possible to achieve in the bilayer system.

All skyrmion bags of the twisted bilayer skyrmion lattices can also be realized in the trilayer system using a mirror-symmetric structure with $\phi_{12} = \phi_{23}$. With such an approach, the smallest skyrmion bags are numerically calculated and depicted in Fig. 3. As in the bilayer system, a change of the center of rotation is related to a changed symmetry of the super-moiré pattern and can be used to realize skyrmion bags of additional sizes. The three different rotation centers $P_1$, $P_2$, and $P_3$ are marked in Fig. 3a, b and c.

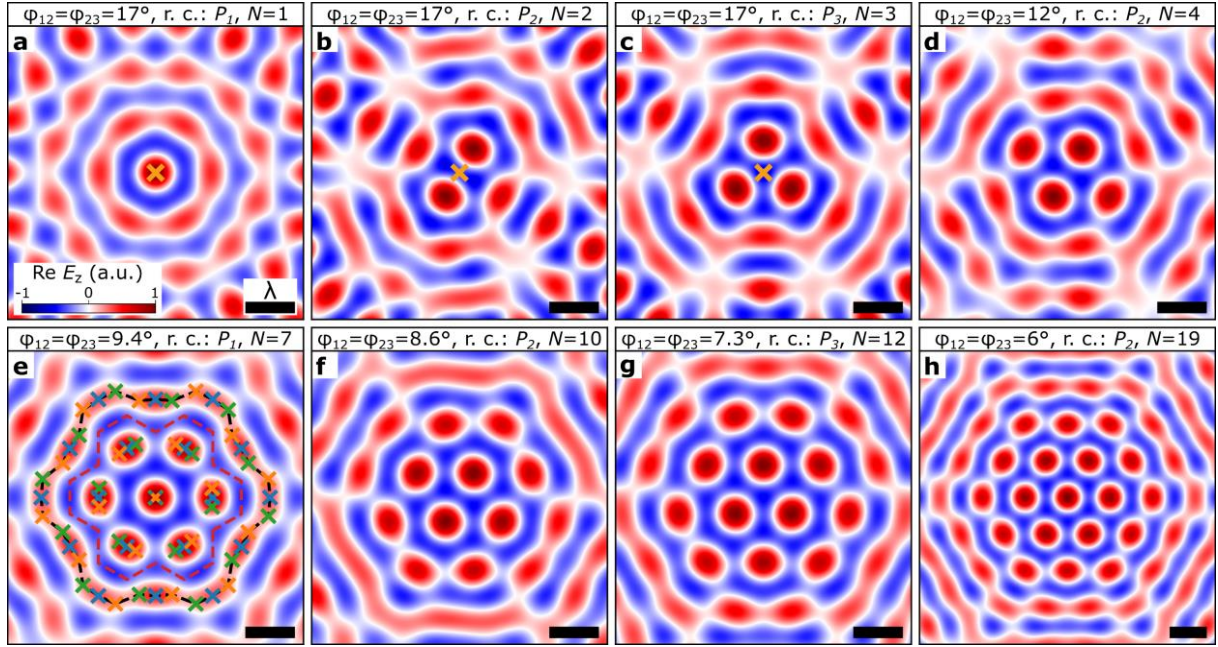

**Figure 3 | Skyrmion bags in super-moiré skyrmion lattices comprising (a) 1, (b) 2, (c) 3, (d) 4, (e) 7, (f) 10, (g) 12 and (h) 19 skyrmions.** Skyrmion bags in twisted trilayer skyrmion lattices with mirror-symmetric twist angles $\phi_{12} = \phi_{23}$. The size of the skyrmion bags is controlled by these twist angles and the rotation centers (r. c.). The three rotation centers $P_1$, $P_2$ and $P_3$ are indicated by the orange markers in a, b, and c, respectively. In e,

the position of the skyrmions of the three skyrmion lattices are indicated by the markers and the integration boundary of the skyrmion bag (the cluster within the bag) is indicated by the black (red) dashed line.

**Robustness of Skyrmion Bags in twisted bilayer and trilayer skyrmion lattices**

The skyrmion bags of Fig. 3 are generated by the interference of three skyrmion lattices instead of two as in Ref. [39]. This is illustrated by the markers in Fig. 3e that indicate the positions of the skyrmions of the individual lattices. It can be recognized that the skyrmions of the cluster within the bag are created by three instead of two overlapping skyrmions that form a more elongated single skyrmion. Additionally, the surrounding bag skyrmion with $S = -1$, is created by the interference of $3/2$ times more skyrmions and is therefore more homogeneous than in twisted bilayer skyrmion lattices. In the following, we illustrate that this increased overlap improves the stability of some skyrmion bags, enabling the choice of the ideal coupling structure to achieve a certain topological texture with maximum robustness.

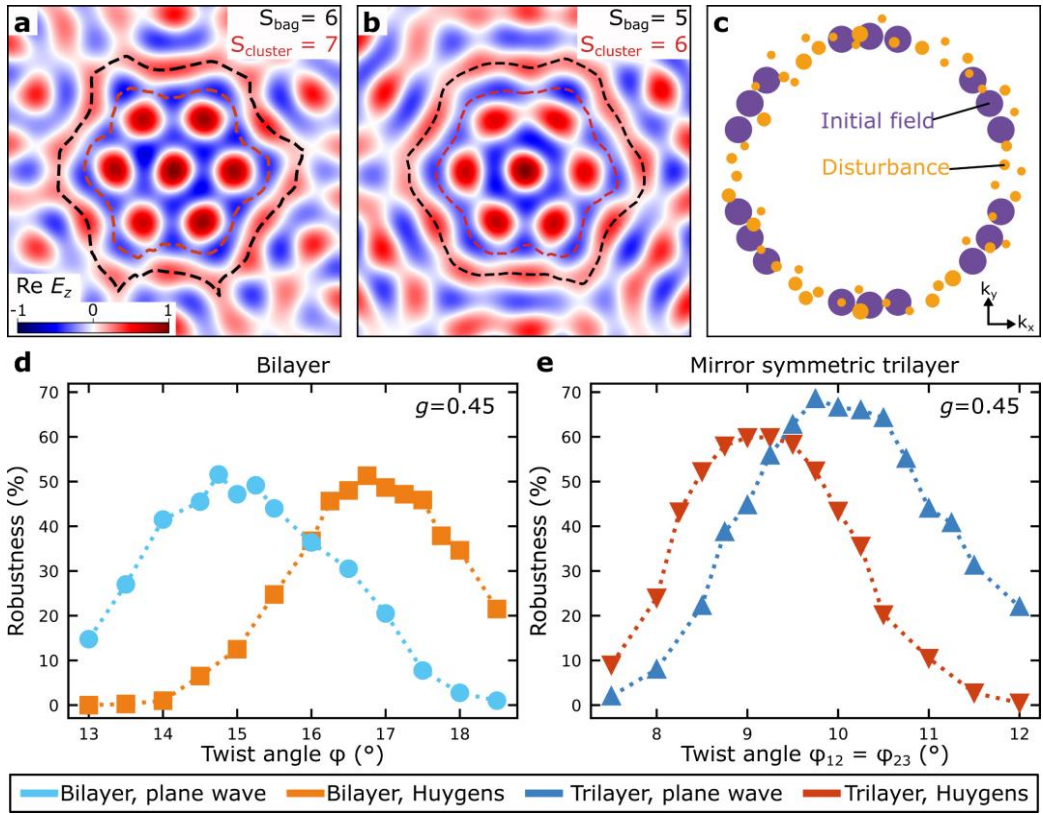

**Figure 4 | Skyrmion bag robustness. a, b** Disturbed electric field distribution of the skyrmion bag with N=7 (see Fig. 3e). The disturbed field in (a) has the correct (preserved) topology and the disturbed field of (b) has an incorrect (changed) topology. **c**, Exemplary Fourier decomposition of the disturbed skyrmion bag. Initial and disturbed field components are colored in purple and yellow, respectively. **d, e**, Robustness of the skyrmion bag with $N = 7$ in twisted (d) bilayer and (e) trilayer skyrmion lattices against perturbations in the form of c. The robustness is defined as the proportion of fields that preserve their topology when the random disturbance is applied. The robustness is presented for initial fields that are calculated using the Huygens approach and in the plane wave approximation.

We quantify the robustness of a topological texture by its ability to preserve its topological numbers when a disturbance field is applied. Here, the disturbance field consists of 50 SPP plane waves with randomly distributed amplitude, phase, propagation constant, and propagation direction (see supplement for details). It is important to note that the propagation constants are chosen to be normally distributed around the propagation constant of the initial field with only a small standard deviation. This is because perturbating waves with other propagation constants could be easily removed via Fourier filtering. Using a perturbation of this kind, the total field in the structure is given by $\boldsymbol{E} = \boldsymbol{E_0} + g \frac{Mean(\|\boldsymbol{E_0}\|)}{Mean(\|\boldsymbol{E_V}\|)} \boldsymbol{E_V}$, where $g$ is a dimensionless parameter that scales the perturbational field $\boldsymbol{E_V}$ in relation to an initial field $\boldsymbol{E_0}$.

First, we apply the disturbance with $g = 0.45$ on the field that is created by bilayer and trilayer structures of a skyrmion bag with $N = 7$ (see Fig. 3e). Two exemplary disturbed vector fields are illustrated in Fig. 4 and b. A schematic Fourier decomposition is depicted in Fig. 4c, where contributions of the initial and disturbance field are colored in purple and orange, respectively. The perturbated field of Fig. 4a represents a deformed skyrmion bag with the same topological numbers of the full skyrmion bag $S_{bag}$ and the skyrmion cluster within the bag $S_{cluster}$. For the derivation of these topological numbers, the integration boundary of the skyrmion number density is set along the black and red dashed lines of Fig. 4a which follow the maximum and minimum out-of-plane field components, respectively. The unchanged topology implies that the perturbated field of Fig. 4a can be created by a continuous deformation of space from the skyrmion bag of Fig. 3e. This is not the case for the perturbated field of Fig. 4b which is why this field distribution possesses different topological numbers. Visually the difference can be identified by noticing that two of the cluster skyrmions have merged. This changes the topological numbers to $S_{cluster} = 6$ and $S_{bag} = 5$. Using perturbational fields with other random parameters, the topology might also be changed by skyrmions of the cluster that are connected to the surrounding bag skyrmion, by a splitting of the skyrmions that otherwise overlap to create the bag skyrmion, or by flipped in-plane field components.

With the help of such random disturbance fields, we quantify the robustness of skyrmion bags by the proportion that preserves its topological numbers $S_{bag}$ and $S_{cluster}$ after application of the random perturbation. Using this approach, we can compare the robustness of skyrmion bags in twisted bilayer and mirror symmetric trilayer skyrmion lattices. The robustness is plotted in Fig. 4d and e as a function of the twist angles. This analysis is conducted for initial fields that are calculated using two different methods. In the plane wave approach, the field is calculated using simple interference of SPP plane waves in the limit of infinitely large coupling structures, while in the other case the field is calculated for the exact coupling structure (as in Fig. 1b), also considering damping of the SPPs (see supplement for details). We note that the comparison between Huygens simulations and plane wave simulations is also important when extending the skyrmion concept from plasmonics to water waves. While the Huygens simulation suits well the plasmonics experiments, the plane wave approximation might be well suited for water waves due to less damping, hence longer propagation lengths and much larger boundary structures from where the waves are emitted.

From the results of Fig. 4d and e it can be concluded that the trilayer skyrmion bag of size $N = 7$ is slightly more robust than the bilayer skyrmion bag against field disturbances that are considered in this approach. In particular, the results show an increase in the robustness from $(52 \pm 1)\%$ to $(60 \pm 1)\%$ for the exact SPP structure and to $(68.5 \pm 1)\%$ for the plane wave interference. The robustness of the exact structure is slightly lower than in the plane wave approach. This is a result of field deviations of the initial field that are attributed to the smaller length of the individual boundaries. In addition to the optimum number of boundaries, we can also derive the ideal twist angles to create the skyrmion bag with optimized robustness which is around 16.75° in the bilayer and 9.25° in the trilayer case.

    In the supplementary information we present the resulting robustness of skyrmion bags comprising 1, 2, 4, 10, 12, 14, and 19 skyrmions as a function of their twist angles. From these results we obtain the ideal twist angles of the bilayer and trilayer structures (see Tab. S1). Using these ideal angles, it is now possible to compare the robustness of the different skyrmion bags.

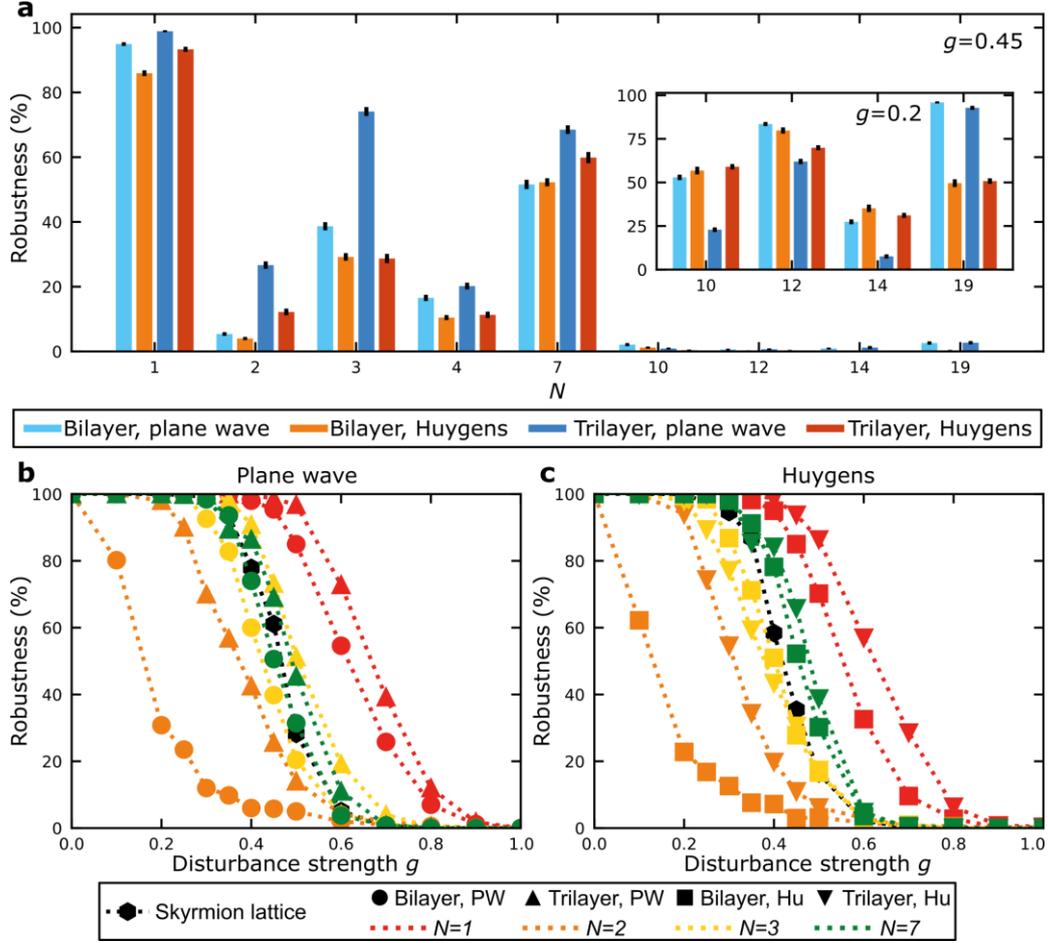

**Figure 5 | Comparison of skyrmion bag robustness. a**, Comparison of the robustness of all different skyrmion bags of Fig. 3. In the inset, the robustness is presented for perturbational fields with lower amplitude to compare the robustness of topologically more unstable skyrmion bags. **b**, **c** Robustness of skyrmion bags and a skyrmion lattice of seven skyrmions as a function of the disturbance strength $g$. The results are shown for initial fields that are calculated using the plane wave approximation (b) and the Huygens approach (c). Bilayer and trilayer coupling structures are distinguished by the markers, while the colors correspond to the size of the skyrmion bags.

The results are illustrated in Fig. 5 for the plane wave and Huygens wave calculation of the initial field. In general, it is noticed that smaller skyrmion bags of the same rotation center have a higher degree of topological stability that stretches over a larger range of twist angles. For different rotation centers, the topological stability varies, which is noticed directly when comparing the robustness of skyrmion bags with $N = 1, 2,$ and $3$. Additionally, smaller skyrmion bags have a larger robustness for trilayer structures, while bilayer structures show higher robustness for larger skyrmion bags. Using a trilayer structure the displacement of skyrmions is increased, which has two effects. The surrounding bag skyrmion with $S = -1$ becomes more homogeneous, which is beneficial for the topological robustness. However, the cluster skyrmions become more elongated, which leads to a more likely overlap of the cluster skyrmions, especially for large skyrmion bags. In total this leads to trilayer structures being superior for small skyrmion bags and inferior for large skyrmion bags.

In Fig. 5b and c the robustness of skyrmion bags with $N = 1, 2, 3,$ and $7$ are shown as a function of the disturbance strength $g$. For weak disturbances, the field textures exhibit a high robustness close to 1, which underlies the stability of the topology in this regime. At a disturbance strength, which is specific to each topological feature and its coupling structure, the robustness starts to decay and reaches vanishing values within $\Delta g \leq 0.3$. Lastly, we also want to compare the robustness of skyrmion bags to the robustness of seven skyrmions in the center of a skyrmion lattice, which is also displayed in Fig. 5b and c. For $g = 0.45$ the robustness of the skyrmion lattice is calculated to $(35.5 \pm 1)\%$ for the Huygens simulation and $(51.5 \pm 1)\%$ for the plane wave simulation of the initial fields and is therefore comparable to the robustness of a skyrmion bag with $N = 7$. This shows that our method can not only be used to optimize coupling structures but also enables the comparison of different topological features and can therefore be used to assess their chances of a successful experimental realization.

**Tab. 1. Calculated skyrmion numbers in super unit cells of trilayer moiré superlattices.** The commensurate twist angles $\phi_{12}, \phi_{13}$ and $\phi_{23}$ are denoted using the coprime integers $m$ and $r$ as explained in the main text. Counting the number of skyrmions of the individual lattices that are contained in the super unit cell yields $S_{individual\ layers}$. However, in the moiré superlattice the three lattices are superimposed within a single layer, which leads to an interference of the lattices. The resulting overlap of skyrmions decreases the observed skyrmion number of the superlattice unit cell to $S_{unit\ cell}$.

| $\phi_{12}$ | $\phi_{13}$ | $\phi_{23}$ | $S_{individual\ layers}$ | $S_{unit\ cell}$ |
|---|---|---|---|---|
| 21.79 (1,1) | 43.57 | 21.79 (1,1) | 49 | 25 |
| 13.17 (2,1) | 26.35 | 13.17 (2,1) | 361 | 139 |
| 21.79 (1,1) | 26.80 (2,3) | 6.01 | 91 | 31 |
| 13.17 (2,1) | 21.79 (1,1) | 8.61 | 133 | 61 |
| 9.43 (3,1) | 13.17 (2,1) | 3.74 | 703 | 211 |

**Outlook**

Our results are helpful for the measurement of twisted trilayer superlattices and skyrmion bags that are created by interference of plasmonic-like waves. This includes phonon polaritonic surface waves [51], as well as acoustic [52] and water wave systems [53]. The robustness of such topological features is controversially discussed [32]. Our analysis is a first approach to quantitatively analyze and compare the robustness of topological features in such systems. It is straight-forward, versatile, and can also be adapted for application in other systems, where the topological textures are also created by the interference of waves.

**Research funding**

The authors acknowledge support from the ERC (Complexplas, 3DPrintedoptics), DFG (SPP1391 Ultrafast Nanooptics, CRC 1242 "Non-Equilibrium Dynamics of Condensed Matter in the Time Domain" project no. 278162697-SFB 1242), BMBF (Printoptics), BW Stiftung (Spitzenforschung, Opterial), Carl-Zeiss Stiftung. T.J.D. acknowledges support from the MPI Guest Professorship Program and from the DFG (GRK2642) Photonic Quantum Engineers for a Mercator Fellowship.